\documentclass[%
 reprint,
 amsmath,amssymb,
 aps,
prl,
floatfix,
]{revtex4-1}

\usepackage{graphicx}
\usepackage{dcolumn}
\usepackage{bm}
\usepackage{subfigure}
\usepackage{hyperref}
\usepackage{wrapfig}
\usepackage{placeins}
\usepackage{natbib}

\begin{document}

\title{Monte Carlo Simulations of Novel Biaxial Ordering in Systems of Uniaxially Interacting Rod-like Ellipsoids} 

\author{Tushar Kanti Bose}
\email{tkb.tkbose@gmail.com}
\affiliation{Department of Theoretical Physics, Indian Association for the Cultivation of Science, Kolkata 700032, India}

\date{\today}

\begin{abstract}
The minimal ingredient to generate a biaxial liquid crystalline ordering is usually considered to be the strongly biaxial interactions breaking the cylindrical symmetry of the uniaxial molecules. Although there is no fundamental reason to forbid a biaxial ordering of pure uniaxial origin, it has been a long standing problem to find a robust demonstration of such phenomenon in systems of rod-like particles. We report here off-lattice Monte Carlo simulations of some new model systems of polar achiral rodlike ellipsoids which spontaneously exhibit novel biaxial smectic phases of pure uniaxial origin. We show that dipolar interactions can generate different biaxial phases of pure uniaxial origin in systems of cylindrically symmetric Gay-Berne ellipsoids for an \textbf{wide variety of length-to-width ratios}. The systems of ellipsoids with low length-to-width ratios exhibit highly tilted biaxial smectic phases in the presence of central axial dipoles. In case of ellipsoids having high length-to-width ratio, the generation of a biaxial phase requires the presence of two parallel axial terminal dipoles. In addition, the phases also exhibit fascinating ferroelectric or striped ordering of the dipolar ellipsoids.
\end{abstract}

\pacs{Valid PACS appear here}
\maketitle

Liquid crystals exhibit an increasingly rich variety of phases due to the complex interplays of different particle shape and molecular level interactions \cite{b1}. The simplest example of a liquid crystal phase is the uniaxial nematic phase in which the anisotropically shaped molecules exhibit a long range orientational order along a particular direction in the absence of any long range positional order. In some liquid crystal phases, the molecules exhibit additional orientational order along a second macroscopic direction. In these `biaxial' phases, we can define a set of perpendicular macroscopic axes of preferential orientations. It is easy to understand the formation of biaxial phases by biaxial molecules, e.g., plank shaped molecules. In a biaxial phase of plank shaped molecules, the molecules arrange themselves in such a way that their similar axes become parallel to each other \cite{b2}. Common examples of biaxial orderings are found in the smectic class of liquid crystalline phases \cite{b4}. In smectic phases, the molecules arrange themselves in two dimensional layers. There the layer normal already exists as a preferred direction due to the layering. In `orthogonal' biaxial smectic phases, the long axes of the biaxial molecules get oriented along the layer normal and their other similar axes also become parallel to each other. In some smectic phases, the molecules are on the average oriented in a direction tilted with respect to the layer normal. Adding a second preferred direction different from the layer normal leads to biaxiality. These smectic phases are called tilted smectic phases \cite{b5}. The origin of such tilted smectic phases have been mostly attributed to the z-like particle shape \cite{b6} and transverse dipolar interactions \cite{b7,b8}. A z-shaped particle is a biaxially shaped particle. On the other hand, a transverse dipole breaks the cylindrical symmetry of uniaxial particles. So, the origins of `orthogonal' and `tilted' biaxial smectic phases are usually considered to be `biaxial'. Can a biaxial liquid crystal phase be formed due to complete uniaxial origin ? Although there is no fundamental reason to forbid such an ordering, the realization of such phenomenon has been very rare. Most of the predictions on the origin of biaxial phases has been based on biaxially shaped particles \cite{b2,b6,b81} or biaxially interacting particles \cite{b7,b8,b18}. An uniaxial model of central quadrupolar Gay-Berne (GB) molecules exhibited a tilted smectic phase for a particular value of the quadrapole moment \cite{b16}. However, the phase disappeared for other lower or higher values of the quadrapole moments. In another work, it was predicted that an uniaxial model of rod-like molecules having two terminal antiparallel dipoles can exhibit a tilted phase \cite{b17}. However, computer simulations of such a model system exhibited only very weakly tilted smectic phases \cite{b18,b7}. So, a robust demonstration of the biaxial smectic generation of uniaxial origin has been missing. In this letter, we simulate the phase behaviors of different model systems of dipolar rod-like particles to robustly demonstrate that dipolar interactions can generate novel biaxial phases of complete uniaxial origin for wide variety of molecular aspect ratios. Dipolar systems have always been quite challenging in terms of predicting their collective behavior, making common sense rather useless. It is interesting to note that in 1992, Wei and Patey discovered that the dipolar interactions can alone generate uniaxial orientational ordering in systems of spherical particles \cite{b12}. They showed that central dipolar spheres can exhibit a nematic phase with global polarization. Camp and Patey also showed that a system of hard spheres with two axial off-central dipoles can exhibit a nematic phase without global polarization \cite{b13}. 
On the other hand, dipolar interactions have also been found to play major roles in generating different interesting phases in systems of highly anisotropic particles e.g., bilayered smectic phase \cite{b14}, ferroelectric nematic phase \cite{b142}, tilted smectic phase \cite{b7}, tilted columnar phase \cite{b141}, stripe domain smectic phase \cite{b15} etc. So far the influence of dipolar interactions has been mostly studied for either spherical or highly anisotropic particles. How do the dipolar interactions influence the collective behavior of particles having shape anisotropies in between the above two limits ? The answer has been mostly unknown. With a motivation of finding an answer to the above question here we first study the phase behavior of central dipolar GB ellipsoids for different aspect ratios \(\kappa=1.5\), 2 and 2.5. We find that GB ellipsoids having aspect ratios \(\kappa=1.5\), \(2\) exhibit novel tilted smectic phases in the presence of central dipoles. These highly tilted smectic phases are indeed examples of biaxial phases of pure uniaxial origin. Such fascinating phase is not found for sufficiently anisotropic particles (\(\kappa=2.5\)) with central dipoles. Then, we show that biaxial phases of pure uniaxial origin can also be found for highly anisotropic ellipsoids (\(\kappa=3\)) in the presence of two terminal parallel point dipole moments placed equidistant from the center of the ellipsoids. The system exhibits a highly tilted smectic phase without global polarization. The phase instead has an interesting antiferroelectric type arrangement of polarized layers of molecules.
\begin{figure*}
\subfigure[\label{fig:p1a}]{\hspace{-.0cm}\includegraphics[scale=.33]{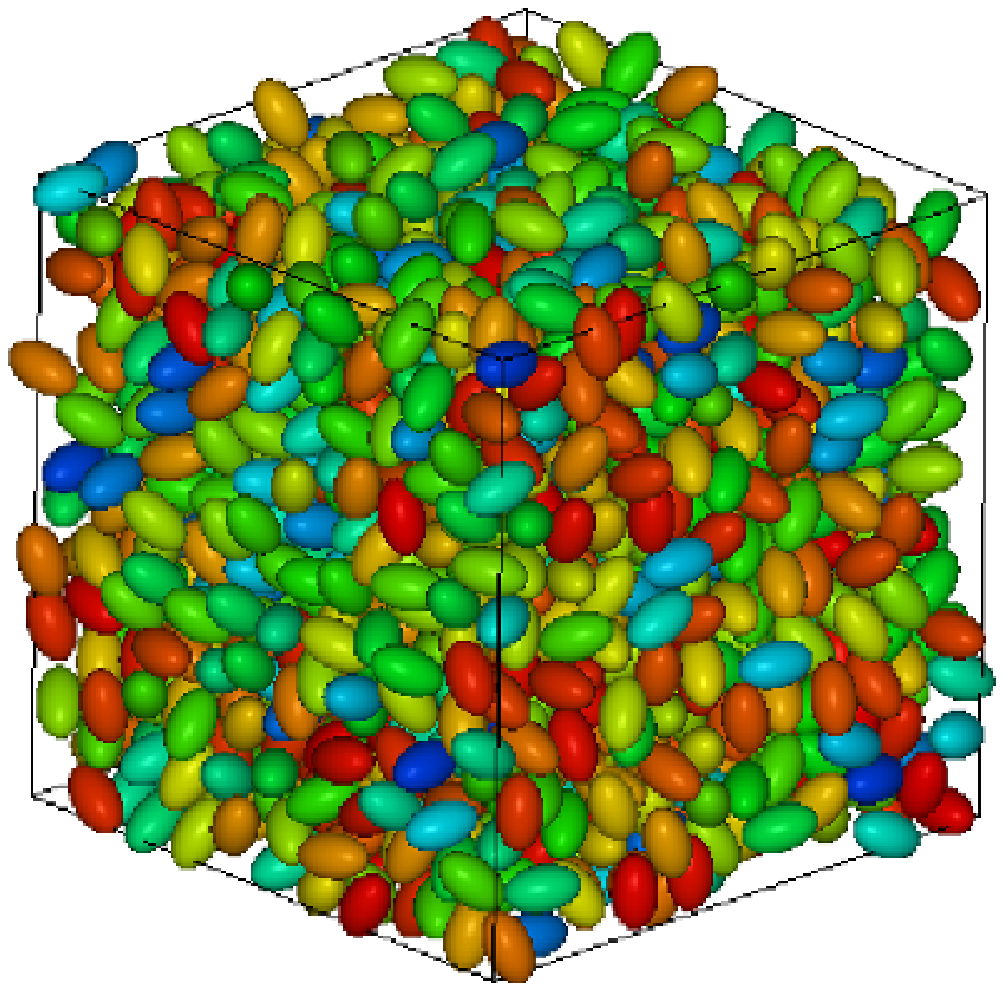}}
\subfigure[\label{fig:p1b}]{\hspace{-.0cm}\includegraphics[scale=.17]{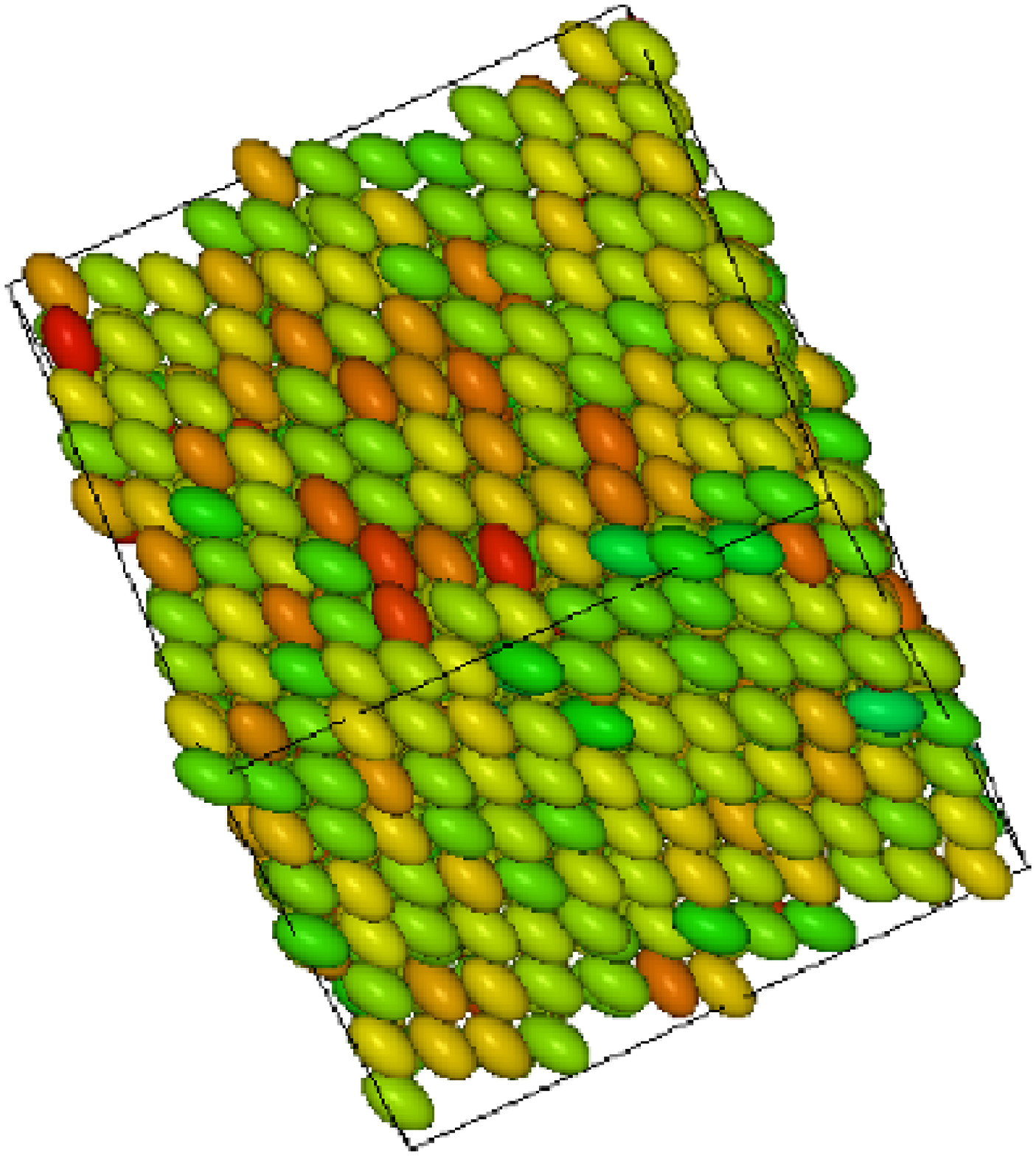}}
\subfigure[\label{fig:p1c}]{\hspace{-.0cm}\includegraphics[scale=.17]{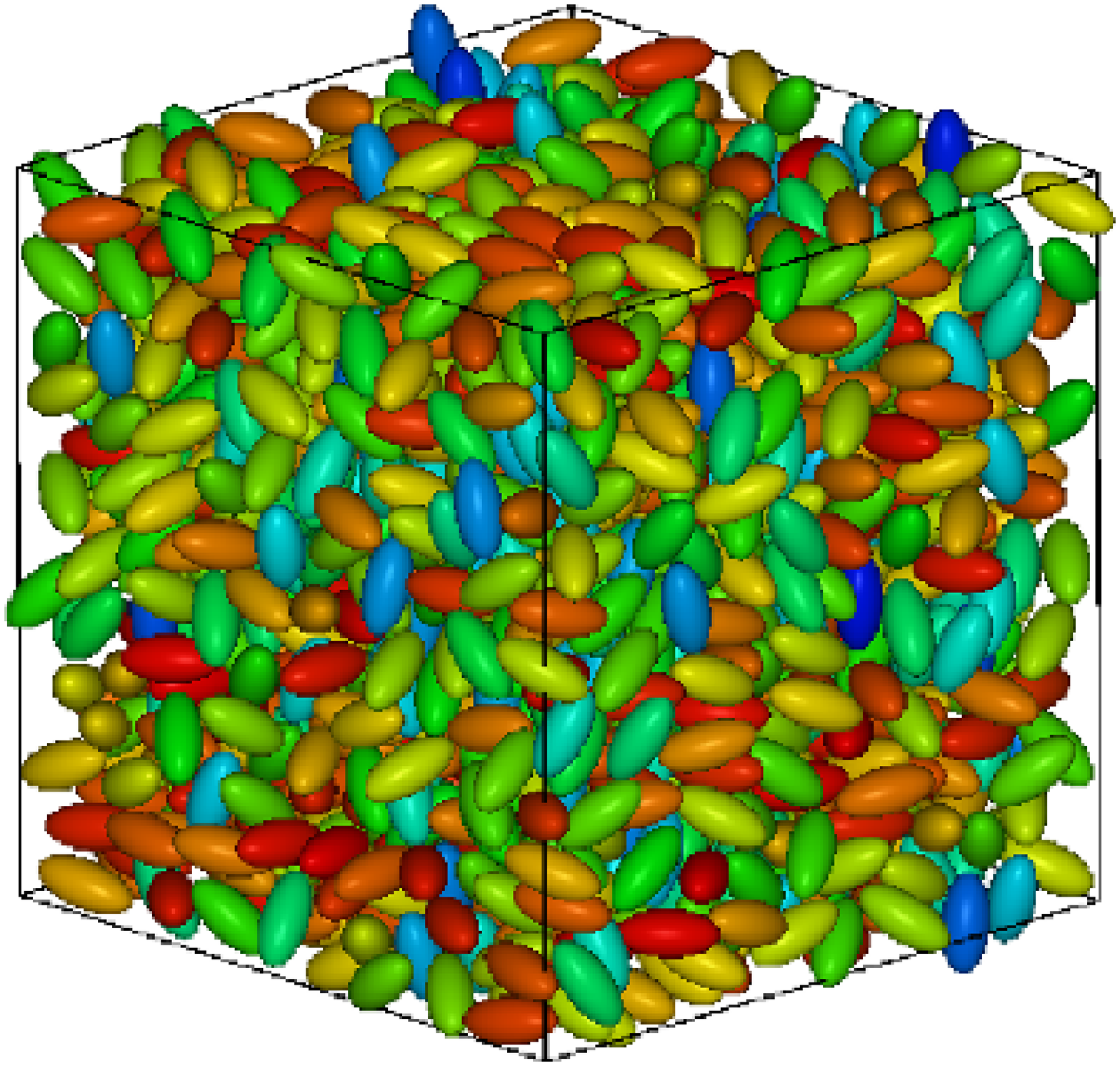}}
\subfigure[\label{fig:p1d}]{\hspace{-.0cm}\includegraphics[scale=.17]{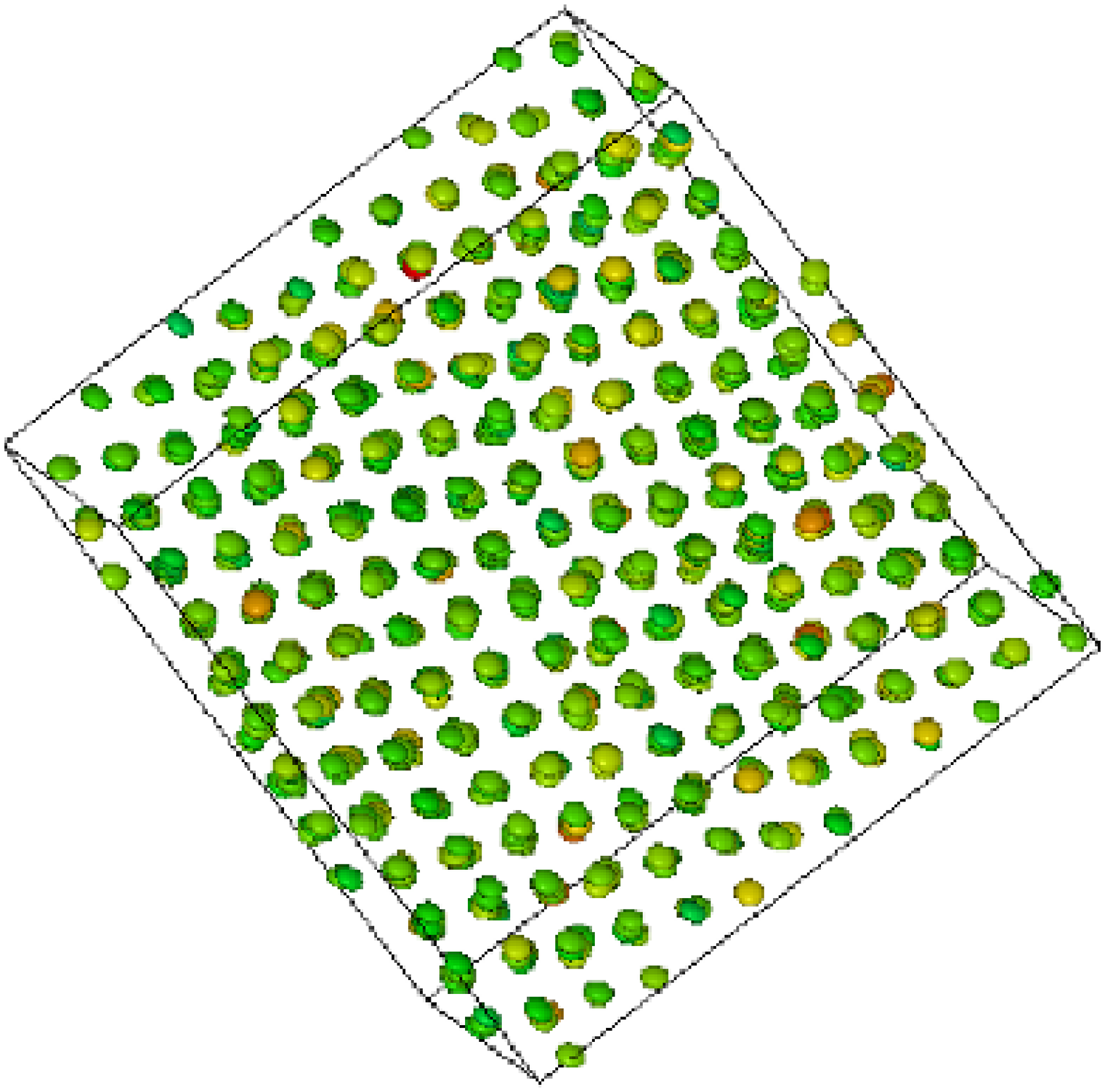}}
\subfigure[\label{fig:p1e}]{\hspace{-.0cm}\includegraphics[scale=.17]{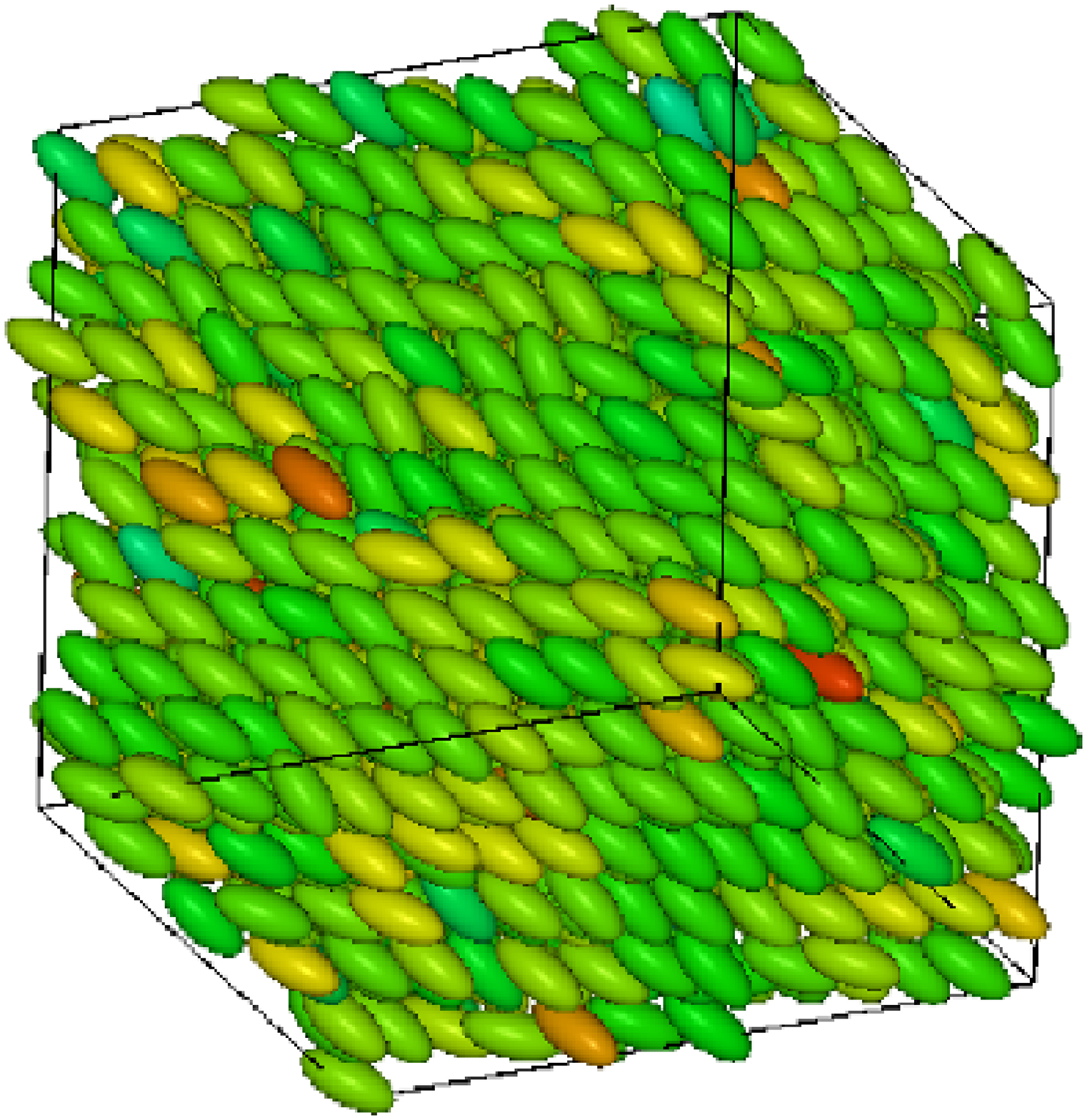}}
\subfigure[\label{fig:p1f}]{\includegraphics[scale=0.9]{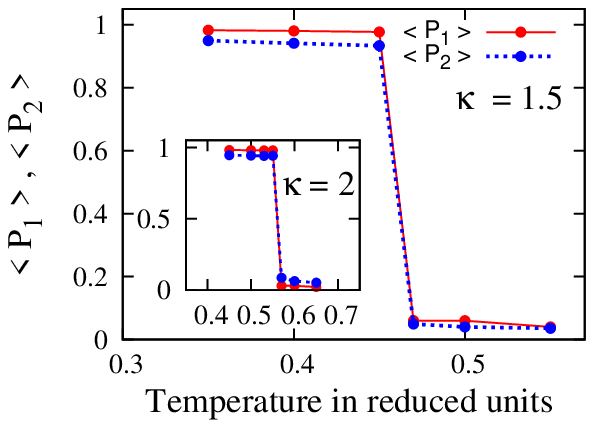}}
\subfigure[\label{fig:p1g}]{\includegraphics[scale=0.6]{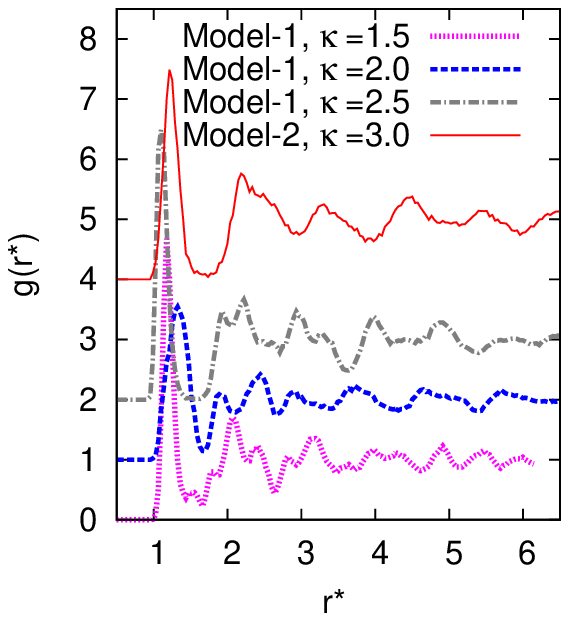}}
\subfigure[\label{fig:p1h}]{\includegraphics[scale=0.85]{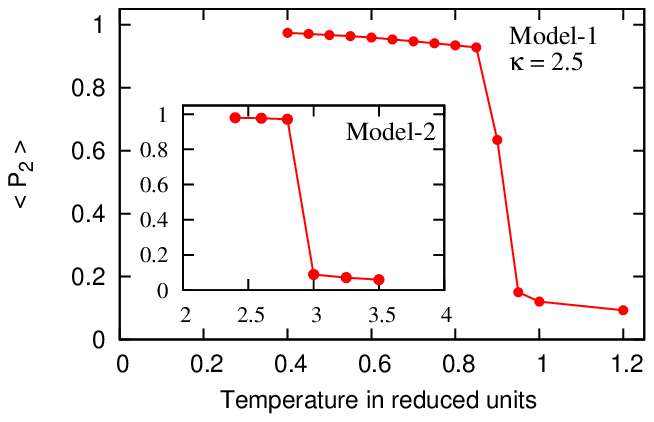}}
\caption{\label{fig:p1}(color online). Snapshots of the liquid crystal phases generated by 1500 `model-1' dipolar ellipsoids at constant pressure \(P^{*}=6.0\)  : (a) Snapshot of the Isotropic phase generated by the ellipsoids with \(\kappa=1.5\) at \(T^{*}=0.55\) (b) Snapshot of the biaxial tilted smectic phase generated by the ellipsoids with \(\kappa=1.5\) at \(T^{*}=0.45\) , (c) Snapshot of the Isotropic phase generated by the ellipsoids with \(\kappa=2\) at \(T^{*}=0.65\) (d) Center of mass positions of the ellipsoids with \(\kappa=2\) in the tilted smectic phase at \(T^{*}=0.53\)(e) Snapshot of the biaxial tilted smectic phase generated by the ellipsoids with \(\kappa=2\) at \(T^{*}=0.53\). Different colors have been used in the above snapshots to indicate different orientations of the particles. The snapshots were generated using the graphics software QMGA \cite{b27}. (f) Variations of the order parameters against temperature for \(\kappa=1.5\) and \(\kappa=2\) (g) The radial distribution functions in the smectic phases for (\(\kappa=1.5,T^{*}=0.45\)),(\(\kappa=2,T^{*}=0.53\)),(\(\kappa=2.5,T^{*}=0.35\)) and (\(\kappa=3.0,T^{*}=2.8\)) as described inside the figure. [ The zero of \(g(r*)\) on the vertical scales
have been shifted for clarity. ] (h) Variation of the nematic order parameter against temperature for \(\kappa=2.5\) and \(\kappa=3 \mbox{ (model-2)}\).} 
\end{figure*}
\begin{figure*}
\subfigure[\label{fig:p2a}]{\hspace{-.0cm}\includegraphics[scale=.22]{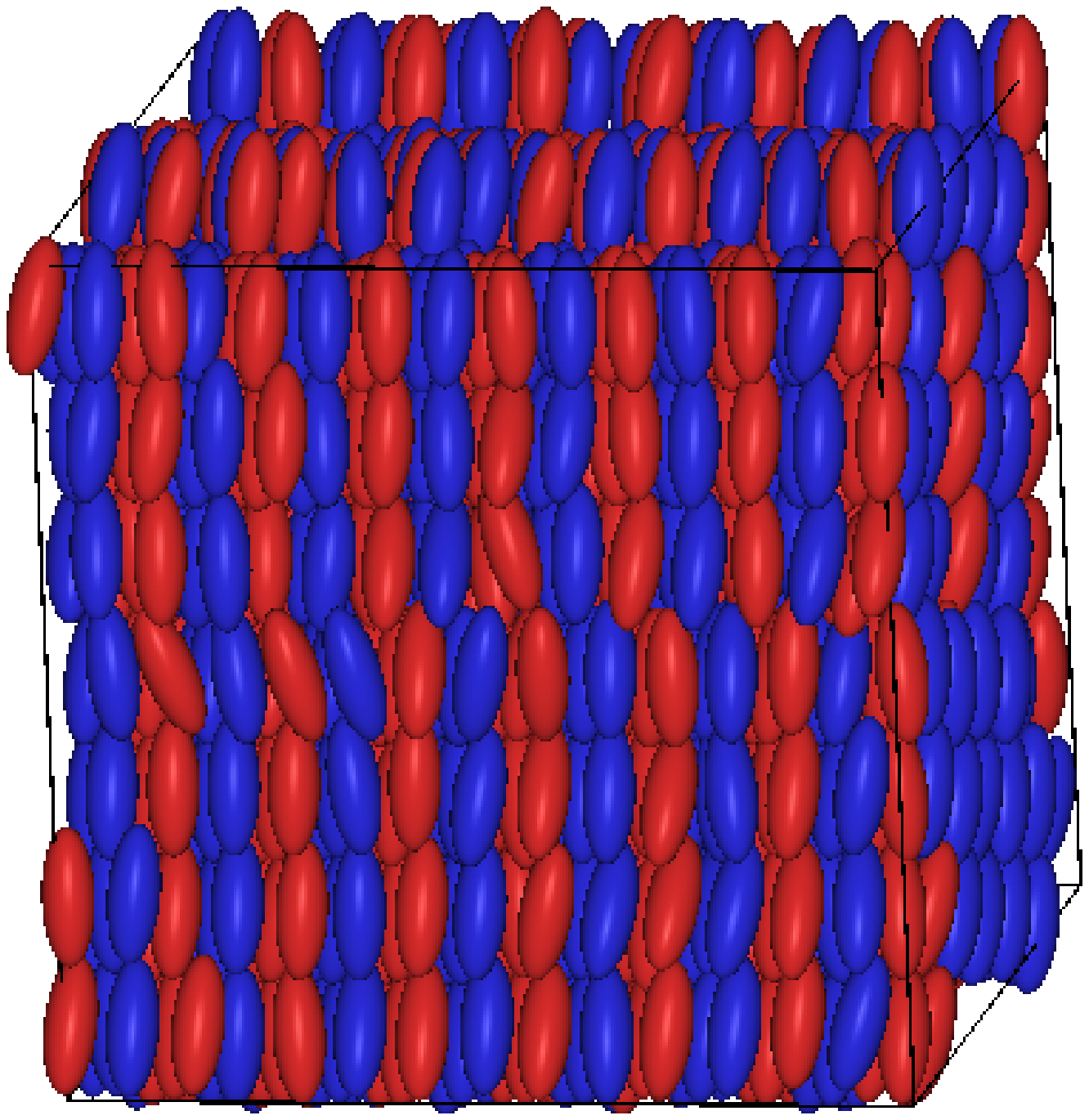}}
\subfigure[\label{fig:p2b}]{\includegraphics[scale=0.280]{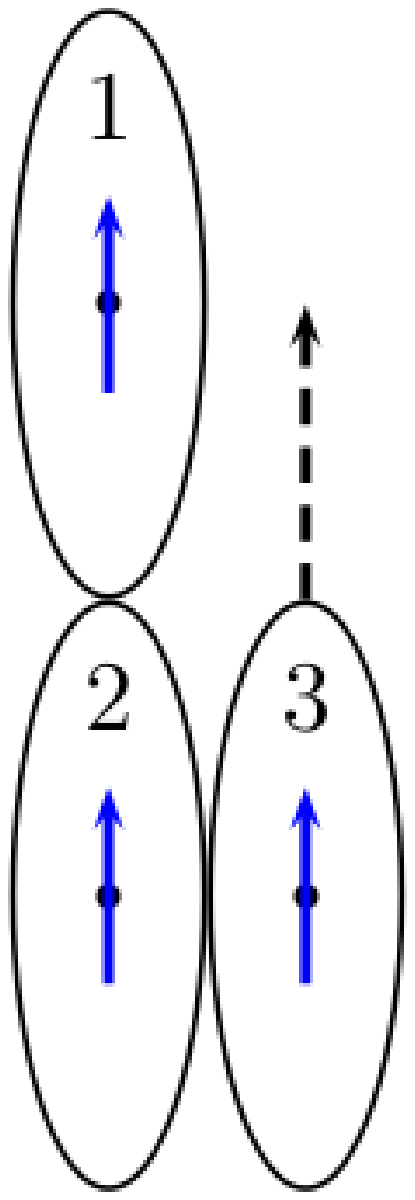}}
\subfigure[\label{fig:p2c}]{\includegraphics[scale=0.280]{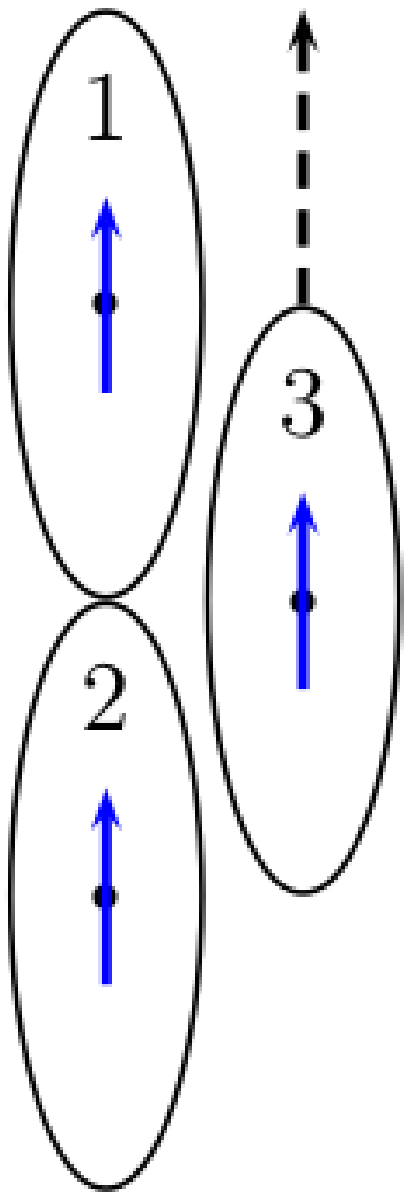}}
\subfigure[\label{fig:p2d}]{\includegraphics[scale=0.5500]{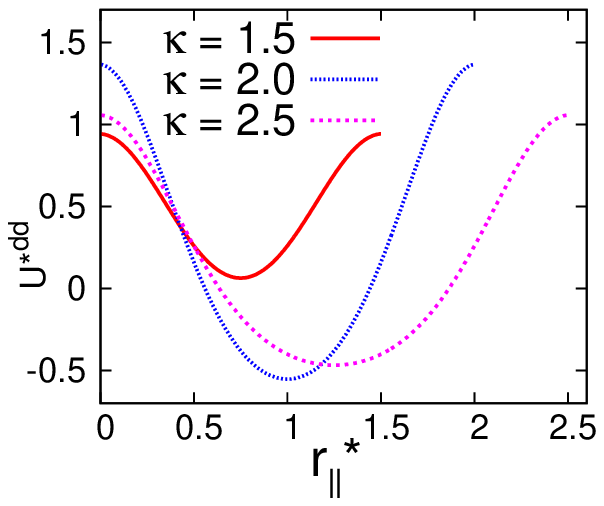}}
\subfigure[\label{fig:p2e}]{\includegraphics[scale=0.5500]{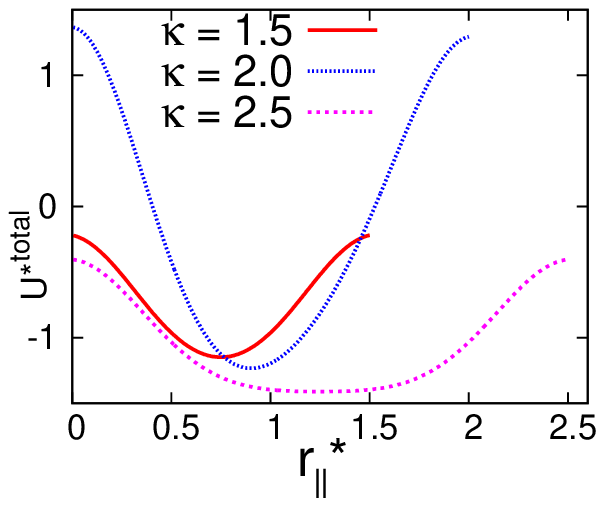}}
\subfigure[\label{fig:p2f}]{\includegraphics[scale=0.15]{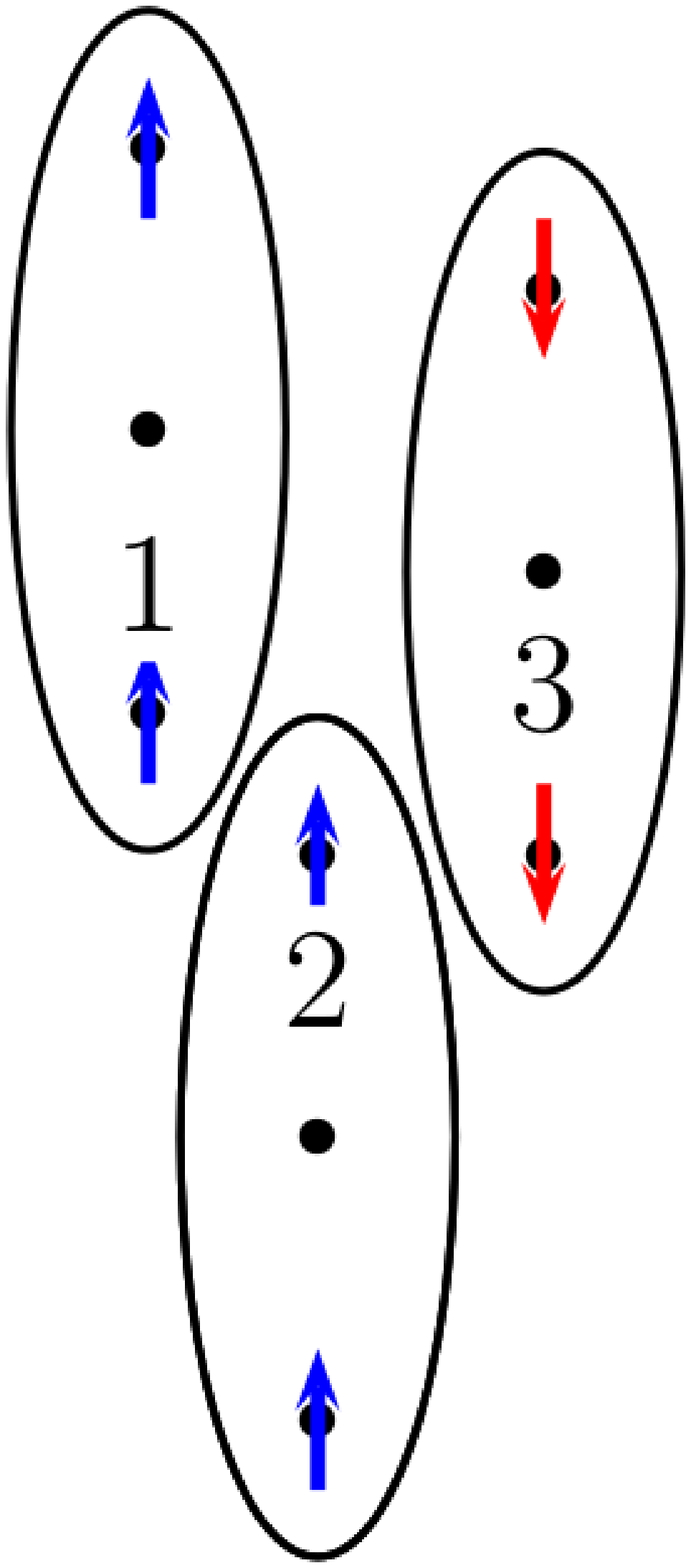}}
\subfigure[\label{fig:p2g}]{\hspace{-.0cm}\includegraphics[scale=.25]{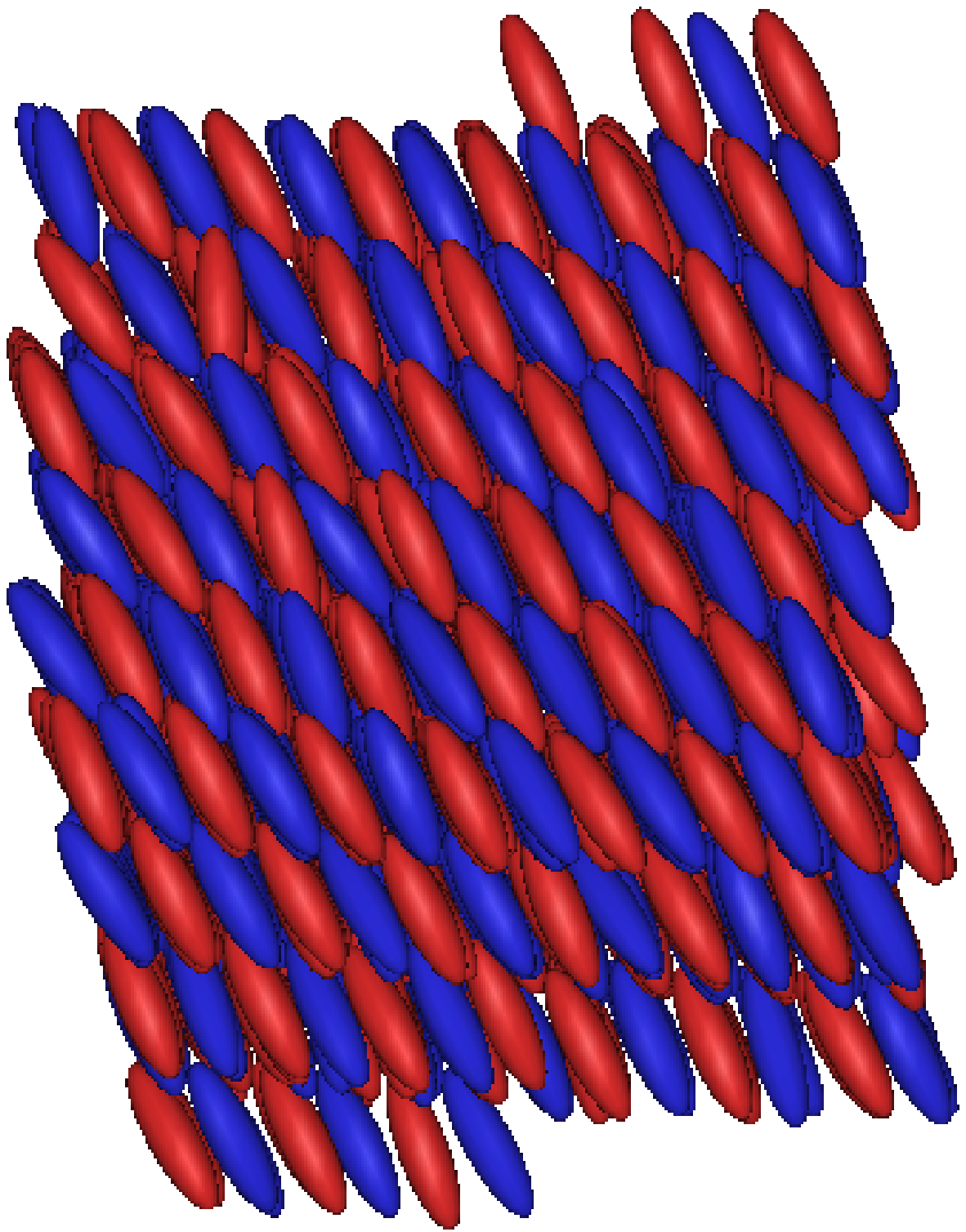}}
\caption{\label{fig:p3}(color online). (a) Snapshot of the smectic phase generated by the `model-1' dipolar ellipsoids with \(\kappa=2.5\) at \((T^{*}=0.35,P^{*}=6.0, N = 1500)\). Two Different colors have been used to indicate two different orientations of the dipoles.(b) A specific configuration of three neighboring `model-1' ellipsoids as described in the text. (c) The specific configuration of three neighboring `model-1' ellipsoids which minimizes the total dipolar interactions as described in the text. (d) Variations of the dipolar interaction energy as a function of parallel component of the pair separation vector. (e) Variations of the total interaction energy as a function of parallel component of the pair separation vector.(f) A specific configuration of `model-2' dipolar ellipsoids. (g) Snapshot of the tilted smectic phase generated by the `model-2' dipolar ellipsoids with \(\kappa=3\) at \((T^{*}=2.8,P^{*}=6.0, N = 1000)\).  }
\end{figure*}
We perform computer simulations of two different model systems of dipolar ellipsoids : model-1 and model-2. In model-1, each ellipsoid is embedded by a central longitudinal dipole. The dipolar ellipsoids are interacting via a pair potential which is a sum of the GB potential \cite{b19} and the electrostatic dipolar interactions. The pair potential between two prolate ellipsoids \textit{i} and \textit{j} is given by \[ \displaystyle{ U_{ij}^{\textrm{GB}}(\mathbf{r}_{ij},\mathbf{\hat{u}}_{i},\mathbf{\hat{u}}_{j})=4\epsilon(\mathbf{\hat{r}}_{ij},\mathbf{\hat{u}}_{i},\mathbf{\hat{u}}_{j})(\rho_{ij}^{-12}-\rho_{ij}^{-6}) } \]
\(\displaystyle{\mbox{ where }\rho_{ij} = ( r_{ij} - \sigma( \mathbf{r}_{ij},\mathbf{\hat{u}}_{i},\mathbf{\hat{u}}_{j} ) + \sigma_{0} ) / \sigma_{0} \mbox{.}}\) Here unit vectors \(\mathbf{\hat{u}}_{i}\mbox{ and }\mathbf{\hat{u}}_{j}\) represent the orientations of the symmetry axes of the molecules, \(\mathbf{r}_{ij}=r_{ij}\mathbf{\hat{r}}_{ij}\) is the separation vector of length \(r_{ij}\) between the centers of mass of the ellipsoids and \(\sigma_{0}\) is the minimum separation between two ellipsoids in a side-by-side configuration determining the diameter of the ellipsoids. The anisotropic contact distance \(\sigma\) and the depth of pair interaction well \(\epsilon\) are dependent on four important parameters \(\kappa,\kappa',\mu,\nu\) as defined in \cite{b20}. Here \(\kappa=\sigma_{e}/\sigma_{0}\) is the aspect ratio of the ellipsoids where \(\sigma_{e}\) is the minimum separation between two ellipsoids in an end-to-end configuration, \(\kappa'=\epsilon_{s}/\epsilon_{e}\) is ratio of interaction well depths in side-by-side and end-to-end configuration of the rod shaped ellipsoids. The other two parameters \(\mu\) and \(\nu\) control the well depth of the potential. \(\sigma_{0}\) and \(\epsilon_{0}\) define the length and energy scales respectively where \(\epsilon_{0}\) is the well depth in the cross configuration. Here, we study the phase behavior of model-1 for the following three values of the aspect ratio \(\kappa\) : 1.5, 2 and 2.5. In case of the other three parameters, we use their original values i.e., \(\kappa'=5, \mu=1, \nu=2\). The corresponding GB interaction cut-off radii for the above mentioned values of \(\kappa\) are taken as 3.5, 4 and 4 respectively in units of \(\sigma_{0}\). The electrostatic interaction energy between two such dipolar ellipsoids is given by \(\displaystyle{ {U_{ij}^{\textrm{dd1}}}=\frac{{\mu}^{2}}{ r_{ij}^{3}}\left[(\bm{\hat{\mu}}_{i}\cdot{\bm{\hat{\mu}}}_{j})-3(\bm{\hat{\mu}}_{i}\cdot\bm{\hat{r}}_{ij})(\bm{\hat{\mu}}_{j}\cdot\bm{\hat{r}}_{ij}) \right]  }  \) , where \(\textbf{r}_{ij} (=\textbf{r}_{j}-\textbf{r}_{i}) \) is the vector joining the two point dipoles \( \boldsymbol{\mu}_{i} \) and \( \boldsymbol{\mu}_{j} \) on the two molecules at the positions \(\textbf{r}_{i} \mbox{ and } \textbf{r}_{j}\). Then the total interaction energy between two dipolar molecules is given by \( U_{ij}^{\textrm{total}}= U_{ij}^{\textrm{GB}}+U_{ij}^{\textrm{dd1}} \). The dipole moment \( \mu^{*}(\equiv \mu /\sqrt{\epsilon_{0}\sigma_{0}^{3}}) = 1.25 \), for a molecular diameter of \(\sigma_{0}= 10\AA\) and an energy term \(\epsilon_{0}=5\times10^{-15}\mbox{ erg}\) corresponds to 2.79 D. The long range dipole-dipole interaction energy has been evaluated using the reaction field \cite{b21} method with dipolar cut-off radius \(r^{*}_{RF}\equiv r_{RF}/\sigma_{0} = 5.0\) and conducting boundary conditions with dielectric constant \(\epsilon_{RF}=\infty\) for the system of \(N = 1000 \mbox{ and } 1500\) dipolar molecules. The reaction field technique has been satisfactorily employed in previous simulation studies of ferroelectric phases \cite{b22,b23,b142}. In model-2, each ellipsoid is embedded by two terminal axial parallel point dipole moments. The dipoles are symmetrically positioned on the long axis of the ellipsoid, at equal distances from the center of the ellipsoid. The dipoles are separated by a distance \(d^{*}(\equiv d/\sigma_{0})=2.0\) along the axis. Here we have used the following GB parameters : \(\kappa=3,\kappa'=5, \mu=1, \nu=3\). The GB cut-off radius is 5 \(\sigma_{0}\). The electrostatic interaction energy between two such dipolar ellipsoids is given by \(\displaystyle{ {U_{ij}^{\textrm{dd2}}}=\sum_{\alpha, \beta=1}^{2}\frac{{\mu}^{2}}{ r_{\alpha \beta}^{3}}\left[(\bm{\hat{\mu}}_{i \alpha}\cdot{\bm{\hat{\mu}}}_{j \beta})-3(\bm{\hat{\mu}}_{i \alpha}\cdot\bm{\hat{r}}_{\alpha \beta})(\bm{\hat{\mu}}_{j \beta}\cdot\bm{\hat{r}}_{\alpha \beta}) \right]  }  \) , where \(\textbf{r}_{\alpha \beta} (=\textbf{r}_{j \beta}-\textbf{r}_{i \alpha}) \) is the vector joining the two point dipoles \( \boldsymbol{\mu}_{i\alpha} \) and \( \boldsymbol{\mu}_{j\beta} \) on the molecules \textit{i} and \textit{j} at the positions \(\textbf{r}_{i \alpha} = \textbf{r}_{i}\pm \frac{d}{2}\bm{\hat{x}_{i}}\) and \(\textbf{r}_{j \beta} = \textbf{r}_{j}\pm \frac{d}{2}\bm{\hat{x}}_{j}\). Then the total interaction energy between two dipolar molecule is given by \( U_{ij}^{\textrm{total}}= U_{ij}^{\textrm{GB}}+U_{ij}^{\textrm{dd2}} \). The dipole moments \( \mu^{*} = 2.0 \), for a molecular diameter of \(\sigma_{0}= 10\AA\) and an energy term \(\epsilon_{0}=5\times10^{-15}\mbox{ erg}\) corresponds to 4.47 D. The long range dipole-dipole interaction energy has been evaluated using the reaction field \cite{b21} method with dipolar cut-off radius \(r^{*}_{RF} = 6.0\) and conducting boundary conditions for the system of N = 1000 dipolar molecules. 

Monte Carlo (MC) simulation studies have been performed in the isothermal-isobaric (constant NPT) ensemble with periodic boundary conditions imposed on systems of ellipsoids. We have performed a cooling sequence of simulation runs along an isobar at a fixed pressure \(P^{*}(\equiv P\sigma_{0}^{3}/\epsilon_{0})=6\) . We start the simulation from an well equilibrated isotropic liquid phase in a cubic box. We then reduce the temperature \(T^{*}(\equiv K_{B}T/\epsilon_{0})\) of the system sequentially to explore the phase behavior. At a given temperature, the final equilibrated configuration obtained from the previous higher temperature is used as the starting configuration. The systems are subjected to equilibrium runs of \(\geq\) \( 3 \times 10^{5}\) MC cycles at each state point [\(p^{*},T^{*}\)]. During a MC cycle, each particle is randomly displaced and reoriented following metropolis criteria where the reorientation moves were performed using Barker-Watts technique \cite{b21}. An attempt to change the volume of the cubic box is also performed in each MC cycle. 
The acceptance ratio of the roto-translational moves and volume moves is adjusted to \(\sim 30\%\) and 40\% respectively. To overcome any possibility of locking in a metastable state, the dipolar particles were also allowed to attempt up-down flip moves exchanging particle tip with bottom with a 20\% frequency with respect to the roto-translational MC moves.

The average orientational order of the particles is monitored by the second-rank orientational order parameter \(P_{2}\) defined by the largest eigenvalue of the order tensor \(S_{\alpha\beta}=\frac{1}{N}\sum_{i=1}^{N}\frac{1}{2}(3u_{i\alpha}u_{j\beta}-\delta_{\alpha\beta})\), where \(\alpha,\beta=x,y,z\) are the indices referring to three components of the unit vector \(\hat{\textbf{u}}\) along the orientation of the particles and \(\delta_{\alpha\beta}\) is the Kronecker delta symbol. The eigenvector associated with the largest eigenvalue defines the primary director. The value of \(P_{2}\) is close to zero in the isotropic phase and tends to 1 in the highly ordered phases. The global ferroelectric order is measured by calculating the average polarization per particle \(P_{1}\) defined by \(P_{1}=\frac{1}{N}\sum_{i=1}^{N}\) \( \hat{\mu_{i}}. \hat{{d}} \) where \(\hat{\textbf{d}}\) is the unit vector along the direction of total macroscopic moment \(\textbf{P}=\sum_{i=1}^{N}\boldsymbol{\mu}_{i}\) and N is the number of molecules in the system. Usually, the biaxial order in a system of biaxial molecules is measured using the order parameter \( \langle R_{2,2}^{2}\rangle=\langle\frac{1}{2}(1+\cos^{2}\beta) \cos 2\alpha \cos 2\gamma -\cos\beta \sin2 \alpha \sin2 \gamma \rangle \) as described in \cite{b25}, where \(\alpha,\beta,\gamma \) are the Euler angles giving the orientation of the molecular body set of axes w.r.t. the director set of axes. However, \( \langle R_{2,2}^{2} \rangle\) is non-zero only for biaxial phases of biaxial molecules. Another order parameter  \( \langle R_{2,0}^{2}\rangle=\langle \sqrt{\frac{3}{8}} \sin^{2}\beta \cos 2\alpha  \rangle \) is in principle different from zero for biaxial phases of uniaxial molecules. However, when the orientational ordering of the long axes along the nematic director is quite strong, i.e., \(\sin\beta\sim0\), the order parameter becomes zero even in the presence of high biaxiality \cite{b18}. We have also seen that \( \langle R_{2,0}^{2}\rangle\) remains zero in the tilted phases observed in the present work. We have measured the radial distribution function \(g(r)=\frac{1}{4\pi r^{2}\rho}\langle\delta(r-r_{ij})\rangle_{ij}\mbox{ ,}\)where the average is taken over all the molecular pairs. In order to verify the fluidity of the ferroelectric phases, we calculated the mean square displacement (MSD) as follows : \( \langle R^{2} \rangle_{\tau} = \frac{1}{N} \sum_{i=1}^{N}[ \textbf{r}_{i}(\tau)-\textbf{r}_{i}(0)]^{2} \) , where \(\textbf{r}_{i}(\tau)\) is the position vector of the i th particle after completion of \(\tau\) MC cycles. In the fluid phases, the mean square displacement steadily increases with increasing \(\tau\) indicating fluid behavior. In contrast for solids \(\langle R^{2} \rangle_{\tau}\) becomes constant as \(\tau\) increases.  \\

We now describe the phase behavior of the model-1 dipolar ellipsoids i.e., the systems of GB particles having central longitudinal dipole moments. The systems exhibit direct Isotropic to novel tilted smectic transitions for both \(\kappa=1.5\) and \(\kappa=2.0\). The transitions occur at the temperatures \(T^{*}=0.45\) and \(T^{*}=0.55\) respectively for \(\kappa=1.5\) and \(\kappa=2.0\). The corresponding snapshots of the Isotropic and the special smectic phases are shown in Figures ~\ref{fig:p1a}-\ref{fig:p1e}. It can be clearly seen from the figures that the smectic phases are highly tilted. The related variations of the order parameters \(\langle P_{1}\rangle \) and \(\langle P_{2}\rangle\) against temperature are shown in Fig. \ref{fig:p1f}. The strong transitions are indicated by the simultaneous jumps in \(\langle P_{1}\rangle \) and \(\langle P_{2}\rangle\) values from \(\sim 0.1\) to \(\sim 0.9\). The strong similarity in the phase behavior of the two systems(\(\kappa=1.5\) and \(\kappa=2.0\)) indicates the dominance of the dipolar interactions over the GB interactions. Figure \ref{fig:p1d} shows the arrangement of the molecular center of mass positions in the tilted phase for \(\kappa=2\). It can be seen that the phase exhibits a rectangular positional order in the layer plane. The tilted phase for \(\kappa=1.5\) also shows similar ordering. Since the smectic phases are highly tilted, the average separations between two neighboring ellipsoids placed in the same layer and that between two neighboring ellipsoids placed in two different but neighboring layers are nearly the same. So, it becomes very difficult to identify the molecules belonging to a particular layer using only their pair separations. The conventional numerical procedure to construct separate layers of molecules and then to find the average layer normal as used in \cite{b7,b18}, fails here. Therefore, an estimation of the tilt angle is not possible using the above method. The structures of the ferroelectric tilted phases for \(\kappa=1.5\) and \(\kappa=2.0\) are similar. The variation of the radial distribution function is shown in Fig. \ref{fig:p1g}. We have also studied the variations of MSD as a function of time. The MSD versus time plots indicated fluid behavior for the tilted phases \cite{b28}. The model-1 ellipsoids exhibit completely different phase behavior in case of \(\kappa=2.5\). For \(\kappa=2.5\), the system exhibits an Isotropic to Nematic phase transition at a temperature \(T^{*}=0.90\). Further decrease of temperature results into a smectic phase at \(T^{*}=0.85\). However, the smectic phase does not show any tilt or global polarization. At lower temperatures, polar domains are formed as shown in Fig. \ref{fig:p2a}. The related radial distribution function shown in Fig. \ref{fig:p1g} exhibits a broken second peak which is the signature of the hexagonal order in the layer plane. The variations of \(\langle P_{1}\rangle\) and \(\langle P_{2}\rangle \) against temperature are shown in Fig.\ref{fig:p1h}. Polar domains in smectic phases of terminal dipolar ellipsoids was reported by Zannoni et al. in ref.\cite{b15}. The nature of polar order in the present case is quite different from that observed in \cite{b15}. 

In order to understand the phase behavior described above, let us consider the schematic arrangements of the ellipsoids shown in Figures~\ref{fig:p2b} and \ref{fig:p2c}. In Fig. \ref{fig:p2b}, it is shown that ellipsoids numbered 1 and 2 are in an end-to-end position and the third one is in a side-by-side position with the ellipsoid numbered 2. All the ellipsoids and the dipoles are oriented in the same direction. Now, the third ellipsoid is moved along the direction of the molecular symmetry axes such that the transverse component \( r_{\perp}\) of the pair separation vector \(\mathbf r_{23}\) between the moving ellipsoid and the ellipsoid numbered 2 remains constant at a value \( r_{\perp}=\sigma_{0}\) as shown in Fig.\ref{fig:p2c}. The variation in the electrostatic part of the total interaction energy among the three dipolar ellipsoids is plotted in Fig. \ref{fig:p2d} as a function of the magnitude of the longitudinal component \(r_{\shortparallel}^{*}\) of the same pair separation vector \(\mathbf r_{23}\). It is evident from Fig. \ref{fig:p2d} that the dipolar separation has a dramatic role on the pair potential. The dipolar interaction results in attractive well minima for all \(\kappa\) at \(r_{\shortparallel}^{*}=\frac{\kappa}{2}\). When we consider the total interaction, the well becomes sharpest and strongest for \(\kappa=2\) which strongly helps in generating the tilted smectic phase Fig. \ref{fig:p2e}. For \(\kappa=2.5\) the interaction well becomes flat. In addition, the side-by-side GB interaction becomes stronger for \(\kappa=2.5\). So, the system exhibit a simple non-polarized smectic B phase. So, this gives an intuitive explanation of the phase behavior described above. For higher values of aspect ratios, the central dipolar ellipsoids are not expected to generate the tilted biaxial phases as found in earlier simulation studies.

 We then studied the phase behavior of the model-2 system i.e., the system of dipolar GB ellipsoids where each ellipsoid has two parallel dipoles symmetrically positioned at two terminal positions. The simulations now become more expensive than model-1. So, we report here the results obtained with \(N=1000\). We start cooling the system from an well equilibrated high temperature Isotropic phase at \(T^{*}=3.5\). The variation of the order parameter is shown in the inset of the Fig.~\ref{fig:p1h}. The system exhibits a direct Isotropic to tilted biaxial smectic transition at \(T^{*}=2.8\). Further decrease of temperature upto \(T^{*}=2.4\) increases the orientational ordering but the smectic phase exhibit no global polarization. The biaxial tilted smectic phase has a fascinating structure. It consist of layers of molecules polarized in opposite direction as shown in Fig.~\ref{fig:p2g} . So, we call it a striped tilted smectic phase. Since, in this case, the ellipsoids are elongated enough and the value of the tilt angle is such that we can successfully use the conventional method of finding the layer normal and tilt angle described before. The average tilt angle \(\sim 28^{\circ}\). The broken second peak in the radial distribution function shown in Fig.~\ref{fig:p1g} indicates hexagonal order in the layer planes. The MSD variations indicates stronger fluid behavior in comparison to the tilted phases of model-1 \cite{b28}. Here, the origin of tilt and related biaxiality is completely different from that of the previously studied model-1. The schematic arrangement of the neighboring molecules in this phase is shown in fig.~\ref{fig:p2f}. It can be seen that the neighboring molecules are arranged in such a fashion that their interaction energy can be minimized. The molecules are positioned such that two oppositely oriented dipoles of two neighboring molecules are situated side-by-side. Two mutually parallel dipoles of neighboring ellipsoids are also positioned in such a manner to reduce their pair interactions. So, here, it is the interaction between the neighboring dipoles of different ellipsoids which generates the biaxial tilted smectic phase with striped antiferroelectric type ordering. So, such Biaxial tilted phases may also be generated for higher values of \(\kappa\).\\

The present work exhibits a fascinating and rare collection of biaxial liquid crystal phases of pure uniaxial origin. Some of the systems studied here generate proper ferroelectric smectic phases of dipolar origin even in the absence of any noncentrosymmetric geometry or chirality of the constituent molecules which are usually considered as the necessary elements for realizing a ferroelectric liquid crystal phase. It should also be emphasized that for the shortest molecules, the system remains isotropic in the absence of dipolar interactions \cite{b29} but shows biaxial smectic order with large tilt angle in the presence of dipolar interactions. It shows the dominating role of the dipolar interactions in generating a biaxial order in the absence of biaxiality in particle shape or in the interactions. This is comparable to the idea that dipolar interaction can generate uniaxially ordered phases in systems of spheres. The biaxial smectic phases are generated for a wide range of shape anisotropies. When the dipole is at the center, the weakly anisotropic GB ellipsoids exhibit a highly tilted smectic order with global polarization. It will be interesting future work to understand the role of dipole strength and orientation in these systems. We have also seen that when the central dipole is replaced by two parallel dipoles placed at the terminal positions, a tilted smectic order is generated again with antiferroelectric type ordering in case of longer ellipsoids. It will be interesting to see the variation in the phase behavior as a function of the dipolar separation.\\ \\

\section{\label{sec:level8} ACKNOWLEDGEMENTS } 
The author thanks Dr. J. Saha for many helpful comments and suggestions.

\bibliographystyle{plainnat}.

\bibliography{PRL}

\end{document}